\documentclass[12pt]{spieman}

\usepackage{amsmath,amsfonts,amssymb}
\usepackage{graphicx}
\usepackage[colorlinks=true, allcolors=blue]{hyperref}
\usepackage{CJK}
\usepackage{bm} 
\usepackage{amsmath}
\usepackage{multirow}
\usepackage{siunitx}
\usepackage{tabularray}
\usepackage{xcolor}
\usepackage{graphicx}
\usepackage{subcaption}
\usepackage{lineno}
\usepackage{soul}
\usepackage[most]{tcolorbox}

\title{Fringing analysis and forward modeling of Keck Planet Imager and Characterizer (KPIC) spectra}

\author[a,$\star$]{Katelyn A. Horstman}

\newcommand{\caltech}{Department of Astronomy, California Institute of Technology, Pasadena, CA 91125, USA}
\newcommand{\gps}{Division of Geological \& Planetary Sciences, California Institute of Technology, Pasadena, CA 91125, USA}
\newcommand{\ucsc}{Department of Astronomy \& Astrophysics, University of California, Santa Cruz, CA 95064, USA}
\newcommand{\keck}{W. M. Keck Observatory, 65-1120 Mamalahoa Hwy, Kamuela, HI, USA}
\newcommand{\ucla}{Department of Physics \& Astronomy, 430 Portola Plaza, University of California, Los Angeles, CA 90095, USA}
\newcommand{\jpl}{Jet Propulsion Laboratory, California Institute of Technology, 4800 Oak Grove Dr., Pasadena, CA 91109, USA}
\newcommand{\ucsd}{Department of Astronomy \& Astrophysics, University of California San Diego, La Jolla, CA 92093, USA}

\newcommand{\berkeley}{Department of Astronomy, University of California at Berkeley, CA 94720, USA}

\newcommand{\northwestern}{Center for Interdisciplinary Exploration and Research in Astrophysics (CIERA), Northwestern University, 1800 Sherman, Evanston, IL, 60201, USA}

\author[b]{Jean-Baptiste Ruffio}
\author[c]{Jason J. Wang}
\author[c]{Chih-Chun Hsu}
\author[a]{Ashley~Baker}
\author[d]{Luke~Finnerty}
\author[a]{Jerry W. Xuan}
\author[a]{Daniel Echeverri}
\author[a]{Yinzi Xin}
\author[a,e]{Dimitri~Mawet}
\author[m]{Geoffrey A. Blake}

\affil[a]{\caltech}
\affil[b]{\ucsd}
\affil[c]{\northwestern}
\affil[d]{\ucla}
\affil[e]{\jpl}
\affil[f]{UK Astronomy Technology Centre, Royal Observatory, Edinburgh EH9 3HJ, United Kingdom}
\affil[g]{\berkeley}
\affil[h]{\keck}
\affil[i]{\ucsc}
\affil[j]{\ucsd}
\affil[k]{Physics and Astronomy Department, Pomona College, 333 N. College Way, Claremont, CA 91711, USA}
\affil[l]{Department of Astronomy, The Ohio State University, 100 W 18th Ave, Columbus, OH 43210 USA}
\affil[m]{\gps}
\affil[$^\star$]{NSF Graduate Research Fellow}

\author[e]{Randall~Bartos}

\author[f]{Charlotte~Z.~Bond}

\author[d]{Benjamin~Calvin}

\author[h]{Sylvain~Cetre}

\author[h]{Jacques-Robert Delorme}
\author[h]{Greg~Doppmann}

\author[d]{Michael~P.~Fitzgerald}
\author[a]{Nemanja Jovanovic}
\author[d]{Ronald~Lopez}

\author[i]{Emily~C.~Martin}
\author[i]{Evan~Morris}

\author[a]{Jacklyn~Pezzato}

\author[a,e]{Garreth~Ruane}
\author[b]{Ben~Sappey}

\author[a]{Tobias Schofield}

\author[i]{Andrew~Skemer}

\author[k]{Taylor~Venenciano}

\author[e]{J.~Kent~Wallace}

\author[l]{Ji~Wang}
\author[h]{Peter~Wizinowich}

\begin{document}
\maketitle

\begin{abstract}
The Keck Planet Imager and Characterizer (KPIC) combines high contrast imaging with high resolution spectroscopy (R$\sim$35,000 in K band) to study directly imaged exoplanets and brown dwarfs in unprecedented detail. KPIC aims to spectrally characterize substellar companions through measurements of planetary radial velocities, spins, and atmospheric composition. The dominant source of systematic noise for KPIC was fringing, or oscillations in the spectrum as a function of wavelength. The fringing signal could dominate residuals by up to 10\% of the continuum for high S/N exposures, preventing accurate wavelength calibration, retrieval of atmospheric parameters, and detection of planets with flux ratios less than 1\% of the host star. To combat contamination from fringing, we identified its three unique sources and adopted a physically informed model of Fabry-P\'{e}rot cavities to apply to post-processed data. We find that this strategy can effectively model fringing in observations of A0V/F0V stars, reducing the residual systematics caused by fringing by a factor of 2. Beyond modeling the fringing signal, we wedged two of the transmissive optics internal to KPIC to eliminate two of the three sources of fringing and confirmed the third source as the entrance window to the spectrograph NIRSPEC. When applied to new data taken with the wedged optics, our previous model of the Fabry-P\'{e}rot cavity reduced the amplitude of the residuals by a factor of 10. 
\end{abstract}


\section{Introduction}
\label{sec:intro}
The goal of the Keck Planet Imager and Characterizer (KPIC) is to spectrally characterize exoplanets by measuring their radial velocities, spins, and atmospheric parameters \cite{Wang2021, Xuan2022, Ruffio2023, Finnerty2023, Xuan2024, Horstman2024, Hsu2024, Morris2024, Costes2024, Echeverri2024Science, Zhang2024, DoO2024, Finnerty2024, Sappey2025, Finnerty2025}. KPIC also seeks to develop key technologies to search for biosignatures such as methane, carbon dioxide, oxygen, and ozone, on potentially habitable planets outside our solar system with the next generation of extremely large telescopes \cite{Decadal2021}. Reaching the photon noise limit for KPIC observations is difficult and requires an accurate model of the data that accounts for the spectra of the host star and companion, as well as characterizable systematics, such as atmospheric variability or optical fringing. 

KPIC (R$\sim$35,000 between $\text{1.94-2.49} \mu$m)\cite{Mawet2017,Delorme2021b} is a fiber-fed high-contrast imaging instrument suite that interfaces with the high-dispersion NIRSPEC echelle spectrograph on Keck 2 \cite{McLean1998, Martin2018}. KPIC residuals show systematic fringing, or periodic oscillations in the continuum flux as a function of wavelength. This type of fringing is caused by the detector or transmissive optics acting as a Fabry-P\'{e}rot cavity within the instrument and has long been a source of instrumental noise within NIR spectrographs. Its ubiquity has created the need to identify sources of fringing and develop various mitigation techniques. Some of the first instances of unwanted fringing in NIR spectrographs are seen in the Space Telescope Imaging Spectrograph (STSI) on the Hubble Space Telescope (HST) (NIR R$\sim$500–1,000 between $\text{0.75–1.027} \mu$m) data. HST/STSI suffers from fringing caused by reflections between the front and back surfaces of the CCD chip. To combat the fringing signal, fringe flats, or tungsten flats taken through a small slit to mimic a point source, were taken to model and remove the signal\cite{Goudfrooij1997, Kimble1998, Malumuth2003}. Although examples of fringing have been documented for over 20 years, modern NIR spectrographs continue to battle this source of noise, since fringing becomes more likely when pushing toward longer wavelengths and higher resolution due to the coherence length of light increasing. The Medium Resolution Spectrograph (MRS) on JWST/MIRI (R$\sim$4,000-1,500 between $\text{5-28} \mu$m) also sees fringing caused by its detectors, mitigating the signal through modeling\cite{Argyriou2020, Gasman2023} while the NIR high-resolution spectrograph iSHELL (R$\sim$80,000 between $\text{2.18-2.47} \mu$m) at the NASA Infrared Telescope Facility (IRTF), requires modeling to account for fringing caused by transmissive optics and reflective coatings\cite{Cale2019}. 

For high signal-to-noise (S/N) observations (S/N $>$ 10), the fringing amplitude can reach up to 10\% of the stellar continuum in KPIC \cite{Finnerty2022}, overwhelming the photon noise for a typical observation. This poses problems for fitting stellar spectra used for deriving wavelength solutions, retrieving accurate elemental abundances in exoplanet atmospheres, and detecting faint planets as the fringing signal can overwhelm spectral features on the order of 1\% the stellar continuum, especially if the fringing period is commensurate with spectral patterns in the target atmosphere (e.g., the CO bandhead at $2.3 \mu$m). Several different attempts have been made to mitigate the fringing signatures in KPIC data. For temporal observations of hot Jupiters, Finnerty et al.\cite{Finnerty2023} removed the time-varying fringing signal attributed to the KPIC optics by using PCA analysis. For directly imaged companion observations, Xuan et al.\cite{Xuan2024AB} incorporated a physics-based approach to model the contaminated residuals of their spectra to account for extra systematics, Ruffio et al.\cite{Ruffio2023} and Horstman et al. \cite{Horstman2024} applied a Fourier filter to remove the main frequencies associated with the periodic fringing signal, and Hsu et al. \cite{Hsu2024PDS} used an approach taken from Cale et al. \cite{Cale2019} to model the residual fringing signal. 

In \autoref{sec:fpc_KPIC}, we explain how several transmissive optics in KPIC/NIRSPEC act as Fabry-P\'{e}rot cavities and the properties of each unique optic are related to the transmission of the fringing signal. In \autoref{sec:solutions}, we create a physically informed model of Fabry-P\'{e}rot cavity to mitigate the fringing signal caused by each transmissive optic in KPIC/NIRSPEC. In \autoref{sec:data}, we describe the application and results of applying the model of Fabry-P\'{e}rot cavities to on-sky observations. Finally, in \autoref{sec:wedge} we describe how we altered the physical optics themselves to remove the most difficult to characterize fringing signals and reapply our Fabry-P\'{e}rot cavity model to on-sky observations utilizing the new optics.

\section{Fabry-P\'{e}rot cavities within KPIC/NIRSPEC} 
\label{sec:fpc_KPIC}
\subsection{Fabry-P\'{e}rot cavities}
\label{sec:fabry}

\begin{figure*}
  \centering
  \includegraphics[trim={0cm 0cm 0cm 0cm},clip,width=1\linewidth]{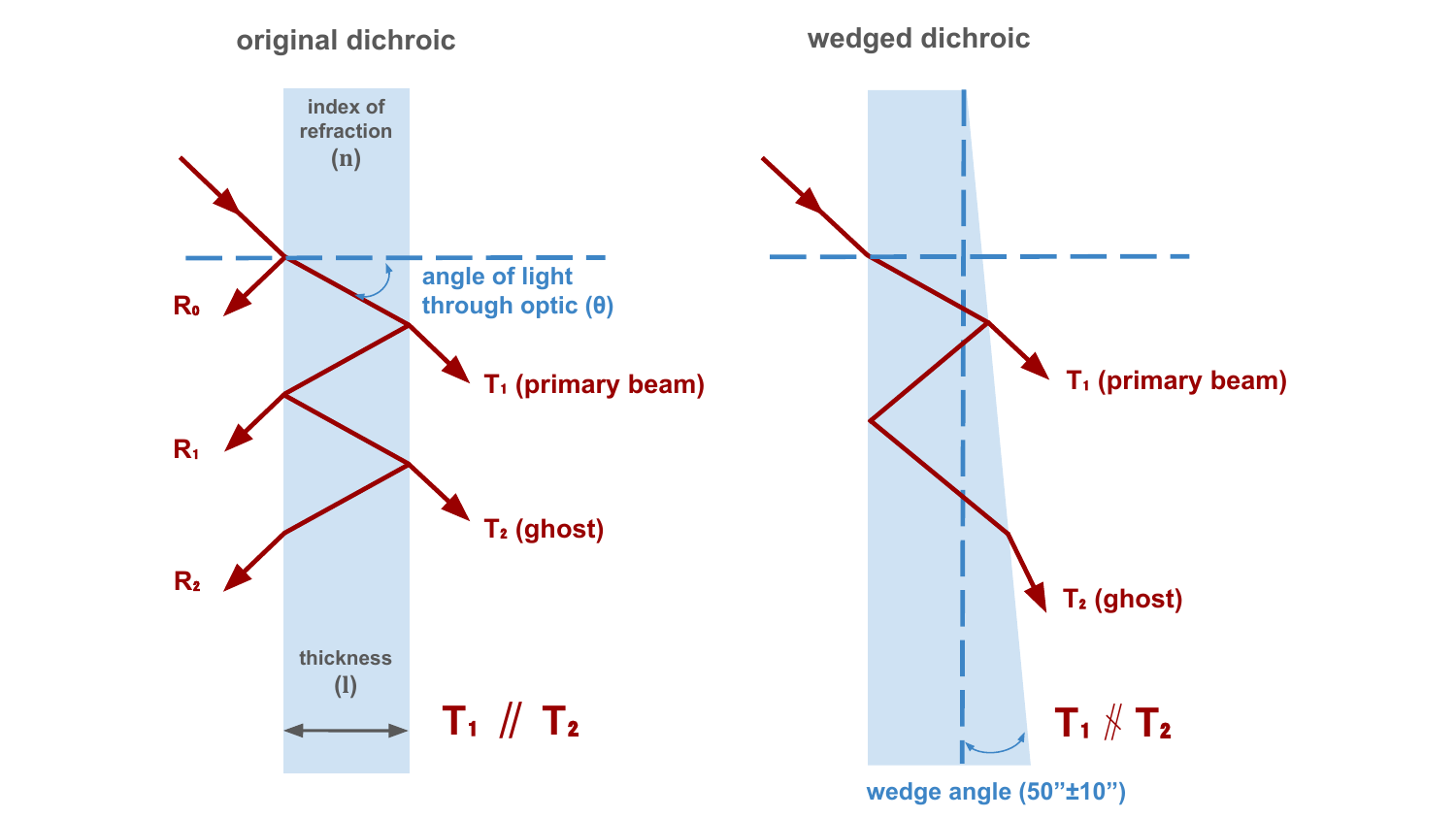}
  \caption{How dichroics can act as Fabry-P\'{e}rot cavities within KPIC/NIRSPEC. \textbf{Left: }The transmissive dichroic within KPIC that acts as a Fabry-P\'{e}rot cavity due to its parallel surfaces. All variables from \autoref{eq:transmission} are included to explain how light propagates through the dichroic and produces fringes. Internal reflections within the dichroic, fringes or ghosts, follow a path parallel to the primary beam. \textbf{Right: }How wedging dichroics can remove unwanted fringing. By introducing a wedge, the internal reflections are offset from the primary beam. The wedging of the dichroics is in explained in further detail in \autoref{sec:wedge}.}
\label{fig:fab_diagram}
\end{figure*}

Fabry-P\'{e}rot cavities are made of parallel interfaces with non-zero reflectivity and can create regular fringes from constructively and destructively interfering light \cite{Perot1899}. The two wave interference equation can be expressed as
\begin{equation}    I=I_{1}+I_{2}+2\sqrt{I_{1}I_{2}}\cos({\phi_{2}-\phi_{1}})
\label{eq:interference}
\end{equation}
where $I_{1}, I_{2}$ are the intensities of each wave while $\phi_{1}, \phi_{2}$ are the respective phases. The total intensity, $I$, of the interfering waves describes an interference pattern that changes with respect to the phase difference between the two waves. The varying transmission function of a Fabry-P\'{e}rot cavity is a direct result of interference between reflections of light and is given by 
\begin{equation}
    T=\frac{(1-R)^{2}}{1-2R\cos\delta(\lambda)+R^{2}}=\frac{1}{1+F\sin^{2}(\frac{\delta(\lambda)}{2})}
\label{eq:transmission}
\end{equation}
where $R$ is the reflectance of the surface of the cavity, $F$ is coefficient of finesse, and $\delta$ is the phase difference. The coefficient of finesse can be rewritten in terms of $R$,
\begin{equation}
    F=\frac{4R}{(1-R)^2}
\end{equation}
while optical path length difference can be expressed as
\begin{equation}
    \delta(\lambda)=\frac{2\pi}{\lambda}2nl\cos\theta
\end{equation}
where $\lambda$ represents the wavelength of light, $n,l$ are the index of refraction and thickness of the cavity, and $\theta$ is the refracted angle relative to the normal of the dichroic surface \cite{Lipson1995}. Therefore, the period only depends on the physical properties of the optic and wavelength of light propagating through the system, as shown in \autoref{fig:fab_diagram}.

\begin{figure*}
  \centering
  \includegraphics[trim={0cm 0cm 0cm 0cm},clip,width=1\linewidth]{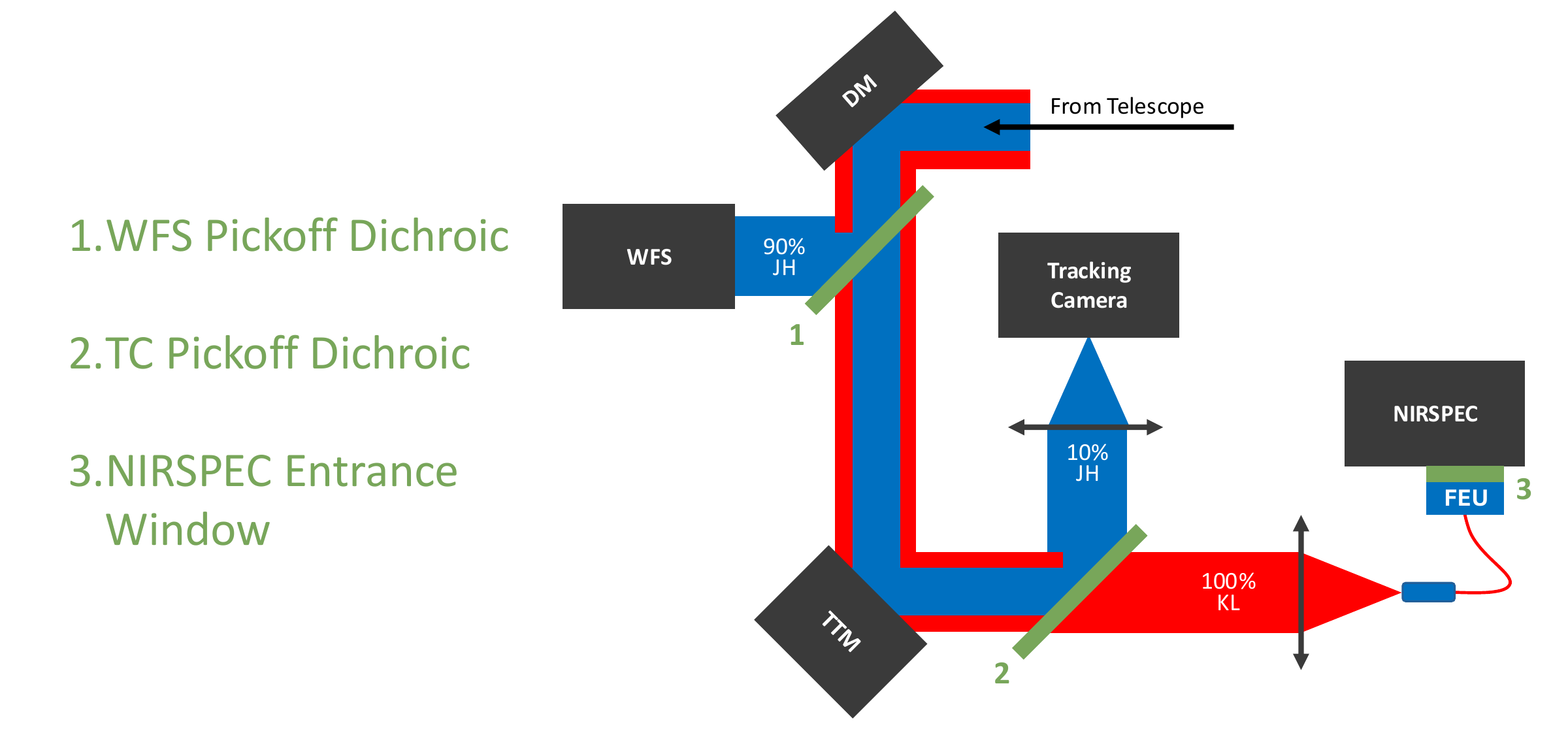}
  \caption{Optical path of KPIC/NIRSPEC. All transmissive optics are represented and labeled in the diagram in green. The following abbreviations are used: Deformable mirror (DM), wavefront sensor (WVS), tip-tilt mirror (TTM), fiber extraction unit (FEU), and tracking camera (TC). The properties of the transmissive optics are explained in \autoref{sec:optics}. Figure adapted from Figure 4a in Echeverri et al.\cite{Echeverri2019}.}
\label{fig:optics}
\end{figure*}

Transmissive optics can unintentionally act like Fabry-P\'{e}rot cavities. When the two surfaces of a transmissive optic are parallel, they can cause light to reflect back into its optical path, forming a Fabry-P\'{e}rot cavity. Interferences, and thus the Fabry-P\'{e}rot effect, appear when the coherence length of the light is larger than the size of the cavity, producing fringes or ghosts. The co-propagating ghost, or faint, reflected light, travels through the system and overlaps with the primary beam. Several transmissive optics used in KPIC and NIRSPEC can act as such cavities, as illustrated in \autoref{fig:optics}.

For 2 \textmu m observations at the resolution of KPIC, the coherence length is around 4 cm for an optic with a refractive index of 1.5. Any unwedged optic thinner than this coherence length could create a Fabry-P\'{e}rot cavity, allowing ghosts to overlap with the primary beam and inject themselves into single mode fibers use by KPIC. In \autoref{sec:optics}, we describe the properties of the transmissive optics in KPIC/NIRSPEC shown in \autoref{fig:optics} that lead to exactly this behavior.
%, which in our case, are the two dichroics and the entrance window of NIRSPEC.

\subsection{Properties of KPIC/NIRSPEC transmissive optics} 
\label{sec:optics}
Fringes within NIRSPAO, or NIRSPEC observations utilizing adaptive optics and nominal slits, have been well documented and are likely a result of the transmissive entrance window of NIRSPEC, affecting both NIRSPAO and KPIC/NIRSPEC science \cite{Brown2002,Drake2005,Blake2010,Hsu2021,Theissen2022}. However, these fringes are not as prominent when the instrument is seeing-limited because of the low spatial coherence of the beam. 
%, and the differing $f$-numbers of the seeing-limited versus AO illumination of NIRSPEC. 
The NIRSPEC entrance window is made of calcium fluoride, which has a refractive index of 1.423 at 2 \textmu m, is 7.5 mm thick, and is not coated with any additional material. The angle of incidence of the beam on the entrance window is 90 degrees. The characteristics of the window are consistent with a 2\text{\AA} period wave, seen in previous NIRSPAO and KPIC/NIRSPEC science observations, when modeled using \autoref{eq:transmission}.

Additionally, KPIC has two dichroics that are transmissive in K and L bands and reflective in J and H bands. One dichroic is designed to send reflected light to a wavefront sensor and the other is designed to send reflected light to a tracking camera, shown in \autoref{fig:optics}. Both dichroics are also made from calcium fluoride and are each 5 mm thick. The angle of incidence of the beam on each dichroic is 45 degrees. The dichroics have a parallelism of less than 2.5$''$, which translates to approximately 5 mas on the sky when taking into account the beam compression factor. The theoretical reflectance of the entrance face is approximately 3.5\%, while the reflectance of the exiting face is estimated to be 2.5\% in K-band. Even though each dichroic has low finesse due to the low reflectivity of the entrance and exit faces, it is sufficient to induce significant fringing since incoherent ghosting scales as $R^{2}$ instead of $2R$ for coherent light. The characteristics of each dichroic replicate a beating fringe pattern with a period between 3-4 \text{\AA}, where the variability of the period is due to a changing optical path length caused by switching the fiber position during observations. Importantly, the dichroics create fringe patterns that {\em combine} with each other, causing high temporal variability of both the fringing amplitude and phase.  

% \section{Solutions to remove unwanted fringing signals}
% \label{sec:solutions}
\section{Physically motivated forward modeling of Fabry-P\'{e}rot cavities} 
\label{sec:solutions}
Assuming the transmissive optics act as Fabry-P\'{e}rot cavities, we incorporate the physically motivated transmission, from \autoref{eq:transmission}, into a forward model of on-sky observations. We model KPIC observations of each target using the python package \texttt{breads}\footnote{\url{https://breads.readthedocs.io/en/latest/}} \cite{Agrawal2023}, following the same methods used in Ruffio et al.\cite{Ruffio2023} and Horstman et al.\cite{Horstman2024}.

We define our forward model as,
\begin{equation}
     {\bm d} = {\bm M}{\bm \phi} + {\bm n},
\end{equation}
where ${\bm d}$ is the data vector of size $N_d$, ${\bm M}$ represents the linear model, ${\bm \phi}$ represents the linear parameters, and ${\bm n}$ is a random vector of the noise with a diagonal covariance matrix ${\bm \Sigma}$, where ${\bm \Sigma}={\bm \Sigma}_0 s^2$. ${\bm \Sigma}_0$ is defined using both the data vector and the standard deviation of the noise, and is multiplied by a free parameter scaling factor $s^2$ to account for any underestimation of the noise.

The linear parameters, ${\bm \phi}$, in our model are the wavelength solution offset, the amplitude of the stellar signal at each node in a $3^{\rm rd}$ order spline model of the stellar continuum, where the number of nodes is fixed. To account for inaccuracies in the continuum of the atmospheric model, ten spline nodes are used in each spectral order ($\Delta \lambda \sim0.05\,\mu$m) for the planet model. This is equivalent to a 200 pixel wide high-pass filter, to balance the number of parameters modeled with the optimal high-pass filter scale of 100 pixels found in Xuan et al.\cite{Xuan2022}. The non-linear parameters in the model ${\bm M}$ include airmass and precipitable water vapor content at the observation site (W.M. Keck Observatory) to inform our telluric model, the radial velocity, $v\sin i$, effective temperature, $\mathrm{log}(g)$, and metallicity of the host to inform our stellar model. We fit all linear and non-linear parameters for every data set.

We introduce eight additional non-linear parameters to account for the fringing based on the physical model of a Fabry-P\'{e}rot cavity:

\begin{equation}
    T=\underbrace{\frac{1}{1+F_{0}\sin^{2}(\frac{a_{0}\lambda+b_{0}}{\lambda})}}_\text{NIRSPEC Entrance Window}\times\underbrace{\frac{1}{1+F_{1}\sin^{2}(\frac{a_{1}\lambda+b_{1}}{\lambda})}}_\text{WFS Pickoff Dichroic}\times\underbrace{\frac{1}{1+F_{1}\sin^{2}(\frac{a_{2}\lambda+b_{2}}{\lambda})}}_\text{TC Pickoff Dichroic}
\label{eq:fringe_model}
\end{equation}

where $F$ is still the coefficient of finesse, but parameters $a$ and $b$ are used to parameterize the phase as a linear function of wavelength. We use $a$ and $b$ to parameterize the optical path length difference because it consists of various parameters dependent on wavelength which can be approximated as a line over a narrow enough wavelength range, such as a single order. Since both dichroics have the same optical properties, we allow them to share a coefficient of finesse, which ultimately dictates the amplitude of the sinusoidal waves.   

\section{Application to data} 
\label{sec:data}
\subsection{Data reduction}
\label{sec:data_reduction}
Detector frames were reduced using the KPIC Data Reduction Pipeline (DRP) \footnote{\url{https://github.com/kpicteam/kpic_pipeline}} following the same procedure as described in Wang et al.\cite{Wang2021}. In summary, the KPIC DRP performs background subtraction, bad pixel correction, and spectral trace calibration to determine the location and width of each of the nine NIRSPEC spectroscopic orders, orders 31-39, on the detector for each of KPIC's four fibers. In general, the most used KPIC order is $2.29-2.34\,\mu$m (order 33), which coincides with the CO bandhead. In our applications of the Fabry-P\'{e}rot cavity model, we focus on fitting order 33 as it is one of the easiest orders to fit and contains important science content. 

As part of the nightly KPIC calibrations, we take spectra of early M giant stars, which have narrow spectral lines due to their slow rotation, to anchor and derive a wavelength solution for each spectroscopic order. The spectral lines from the M-calibrator star and telluric lines from the atmosphere are modeled with a PHOENIX model \cite{Husser2013} and the Planetary Spectrum Generator \cite{Villanueva2018}, respectively, to obtain best fit parameters for the final wavelength solution in each order.

%\subsection{Before wedging dichroics}
The difficulty of using this model lies in the fact that the fringe phase depends on the angle of incidence on the cavity. The fringing caused by NIRSPEC entrance window remains stable over time, the fringing caused by the KPIC dichroics is highly variable since changing the fiber used for observation or offsetting to an off-axis companion changes the angle of incidence and thus phase of the fringes. The distance between fibers, 0.8$''$, roughly corresponds to one-third of a fringe wave. Consequently, developing a single model to account for fringing in every observation is difficult. 

\begin{figure*}
  \centering
  \includegraphics[trim={0cm 0cm 0cm 0cm},clip,width=1\linewidth]{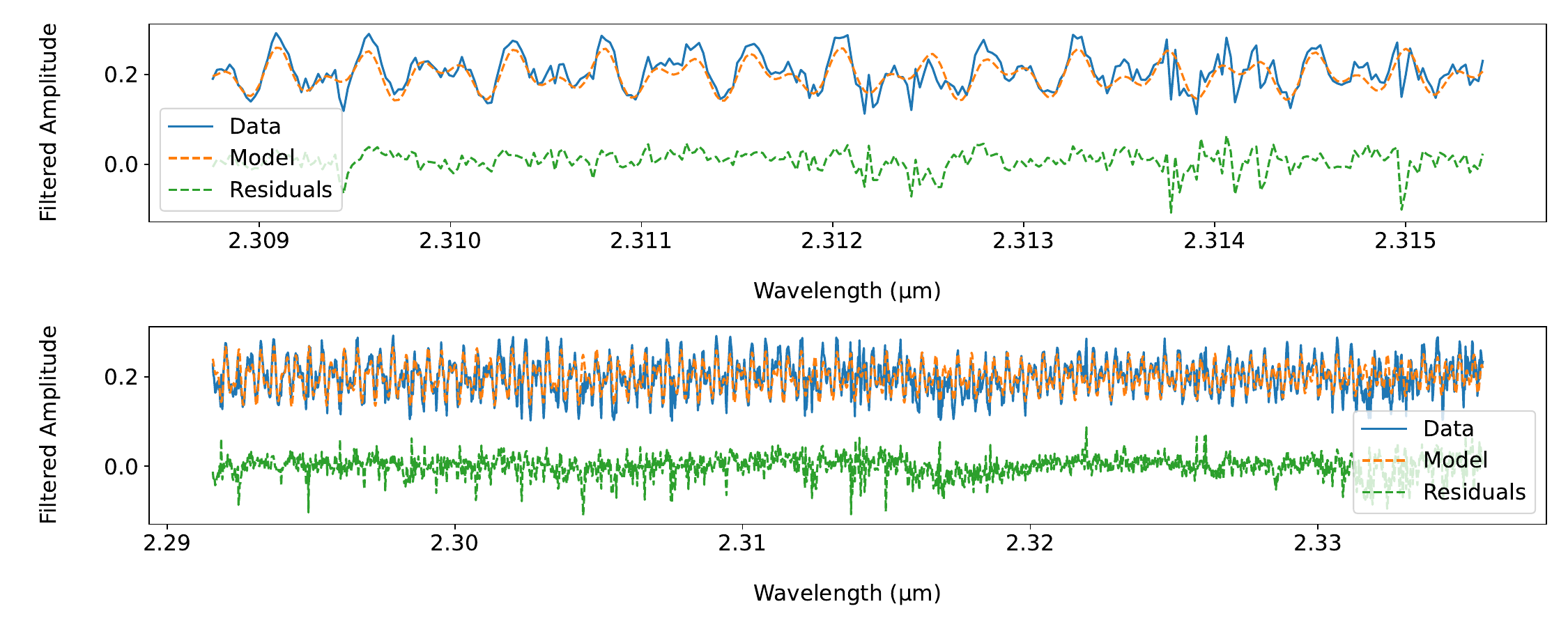}
  \caption{Fringing signal extracted from the A0 calibration star Zeta Aquilae in order 33 ($2.29-2.34\,\mu$m). \textbf{Top:} Zoomed in fringing signal fit to a normalized stellar spectrum using \autoref{eq:fringe_model}. The blue line depicts the data, while the model is in orange and the residuals are in green. This region is free of telluric contamination, so the modulation is exclusively due to fringing. \textbf{Bottom:} Same as top panel, except for the entire order. Best fit parameters can be found in \autoref{table:fringe_fit}.}
\label{fig:fringes}
\end{figure*} 

Thus, instead of developing a singular model for all observations, we create a framework for fitting a model to a variety of observations. First, we extract the fringes from a spectrum with few stellar or telluric lines by dividing our data by the best fit model for a A0V telluric calibration star. We fit our mathematical model, \autoref{eq:fringe_model}, to the extracted empirical fringes to find values for each of the eight parameters. To do this, we implement multiple grid searches over our parameter space to account for a non-convex cost function. The values used to define the grid search are consistent with physical model parameters, matching in expected amplitude and phase, and the residuals are consistent with noise. For $F$, we define a 10,000 step grid with bounds corresponding to the reflectance ranging between 0.1$\%$ and 15$\%$. For $a, b$, we define two, 10,000 step grids with bounds corresponding to refracted angle, $cos\theta$, ranging between 0 and 55 degrees. First, we constrain the individual phases ($a, b$) for each dichroic separately to keep the search computationally feasible. Then, we perform an additional search to constrain the two remaining amplitude terms ($F_0, F_1$). After using the grid search to find a mode of the cost function that is locally convex, we run a Nelder-Mead optimization to refine the best fit parameters. \autoref{fig:fringes} shows the fringe extraction and fitting for an example A0V star, Zeta Aquilae while \autoref{table:fringe_fit} shows the fitting results to each parameter. 

\begin{table}[]
\begin{center}
\begin{tabular}{|l|l|l|l|}
\hline
      & $\hphantom{-}$zet Aql              & $\hphantom{-}$HR 8799            & $\hphantom{-}$HIP 61960 \\
\hline
$F_0$ & $\hphantom{-}5.26\times10^{-2}$    & $\hphantom{-}1.78\times10^{-1}$    &$\hphantom{-}6.36\times10^{-1}$\\
$a_0$ & $-4.8144\times10^{2}$              & $-4.9823\times10^{2}$              &$-1.1150\times10^{3}$\\
$b_0$ & $\hphantom{-}6.8153\times10^{4}$   & $\hphantom{-}6.8194\times10^{4}$   &$\hphantom{-}6.8725\times10^{4}$\\
$F_1$ & $\hphantom{-}5.31\times10^{-2}$    & $\hphantom{-}1.56\times10^{-1}$    &$\hphantom{-}$NA       \\
$a_1$ & $-5.3899\times10^{1}$              & $\hphantom{-}5.7653\times10^{1}$   &$\hphantom{-}$NA       \\
$b_1$ & $\hphantom{-}3.9930\times10^{4}$   & $\hphantom{-}3.9938\times10^{4}$   &$\hphantom{-}$NA       \\
$a_2$ & $-7.8398\times10^{1}$              & $-7.4379\times10^{1}$              &$\hphantom{-}$NA       \\
$b_2$ & $\hphantom{-}3.9999\times10^{4}$   & $\hphantom{-}3.9989\times10^{4}$   &$\hphantom{-}$NA       \\
\hline
\end{tabular}
\end{center}
\caption{Best fit forward modeling parameters to \autoref{eq:fringe_model}. The fit to Zeta Aquilae is to the normalized (data/model) fringing signal, while the fit to HR 8799 and HIP 61960 is to the stellar spectrum, with parameters corresponding to unwedged and wedged dichroics respectively. After wedging the dichroics in April of 2024, we confirm the absence of the fringing signal associated with the KPIC dichroics. We mark the corresponding parameters ($F_1, a_1, a_2, b_1, b_2$) as NA to indicate that we no longer fit for them in the HIP 61960 spectrum. Values for $F$ are consistent with reflectance between 1 and 12 $\%$, while values for $a,b$ are consistent with a refracted angle, $cos\theta$, of $\sim$ 0 degrees for the NIRSPEC entrance window and $\sim$ 25 degrees for both KPIC dichroics.} The parameter $b$ is in units of $\mu m$, while $a$ and $F$ are dimensionless quantities.
\label{table:fringe_fit}
\end{table}

\subsection{Application to F0V host star} 
\label{sec:A0}

To test our model of observations on a host star, we again applied the fringing model to a spectrum with very few stellar lines to evaluate its performance with minimal noise sources. We applied the fringing parameters found in \autoref{table:fringe_fit} for Zeta Aquilae to an on-axis, bright star, HR 8799, as a best guess. We then ran a Nelder-Mead optimization to find the best fit parameters for our HR 8799 observation. 

\begin{figure*}
  \centering
  \includegraphics[width=\linewidth]{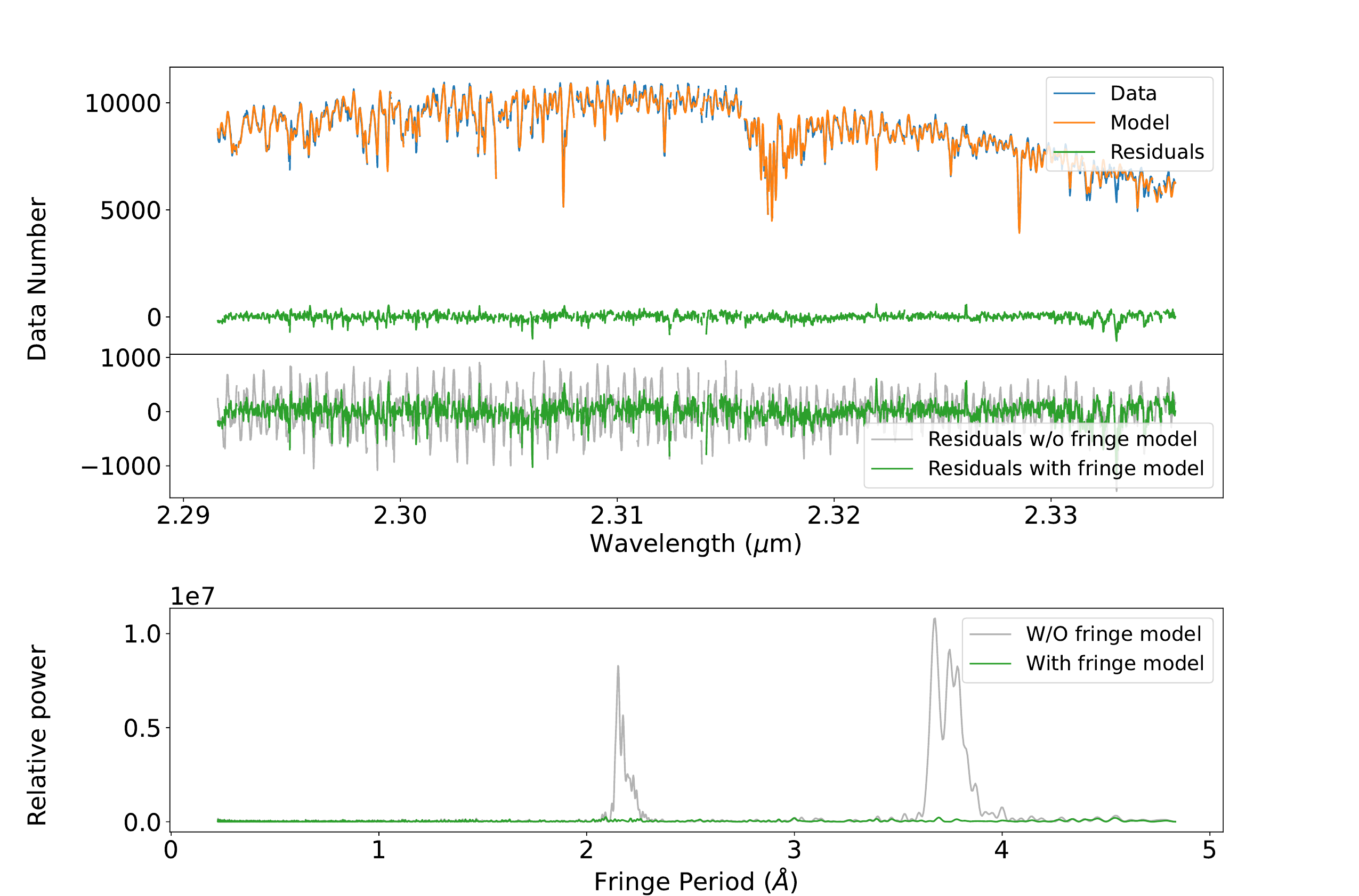}
  \caption{Example of forward modeling of the KPIC fringing signal, before wedging the dichroics, on a telluric calibrator star. \textbf{Top:} Spectrum of HR8799 after incorporating a fringing signal in the forward model framework. The data is in blue, the model is in orange, and the residuals (data-model) are in green. \textbf{Middle:} The residuals of the model including the fringing signal and the model not including the fringing signal. The residuals incorporating the fringing model are in green, while the residuals that do not have any fringing mitigation are in gray. \textbf{Bottom:} A Lomb-scargle periodogram of the residuals of the model including the fringing signal and the model not including the fringing signal. The residuals incorporating the fringing model are in green, while the residuals that do not have any fringing mitigation are in gray. The fringing signal due to transmissive optics within KPIC/NIRSPEC are clearly suppressed at 2\text{\AA} and between 3-4 \text{\AA} by the quantitative modeling framework, reducing the amplitude of the residuals by a factor of 2.}
  \label{fig:HR8799}
\end{figure*}

\autoref{fig:HR8799} shows the fit of the forward model incorporating the fringing signal. The model was able to replicate the fringing pattern from all three sources, suppressing the fringing signal related to both the entrance window and dichroics, reducing the amplitude of the residuals by a factor of 2. The best fit parameters can be found in \autoref{table:fringe_fit}.

\subsection{Application to M-giant wavelength calibration star} 
\label{sec:M}

\begin{figure*}
  \centering
  \includegraphics[width=\linewidth]{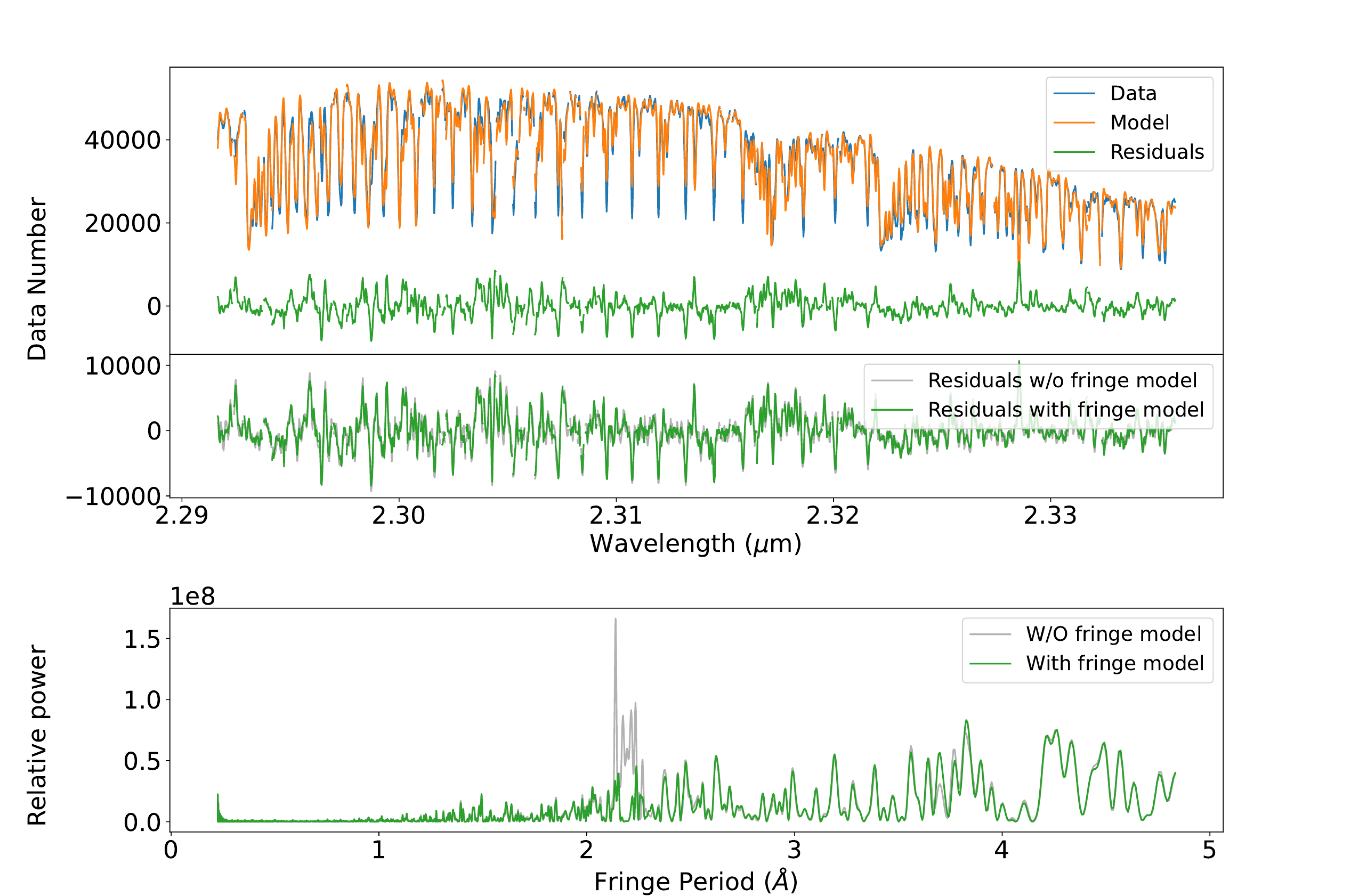}
  \caption{Example of forward modeling the KPIC fringing signal before wedging the dichroics on wavelength calibrator star. \textbf{Top:} Spectrum of HIP 95771 after incorporating a fringing signal in the forward model framework. The data is in blue, the model is in orange, and the residuals (data-model) are in green. There is mismatch between the depth of stellar lines in the model versus in the data. \textbf{Middle:} The residuals of the model including the fringing signal and the model not including the fringing signal. The residuals incorporating the fringing model are in green, while the residuals that do not have any fringing mitigation are in gray. \textbf{Bottom:} A Lomb-scargle periodogram of the residuals of the model including the fringing signal and the model not including the fringing signal. The residuals incorporating the fringing model are in green, while the residuals that do not have any fringing mitigation are in gray. Although the fringing signal appears to be slightly suppressed, the mismatch between the stellar lines of the model and data are on the same order of magnitude as the fringing signal. The power spectrum of the stellar atmosphere and that of the fringing signal have significant overlap, making it much more difficult to fit the fringes.}
  \label{fig:HIP95771}
\end{figure*}

Next, we applied the fringing model to an M-giant star, primarily used for calibrating wavelength solution for each order in KPIC due to the star's deep, abundant stellar lines. Modeling the fringing signal for wavelength calibration stars is important for both obtaining a more precise wavelength solution and exploring whether the wavelength solution is biased by the fringes. \autoref{fig:HIP95771} shows the fit of the forward model incorporating the fringing signal.

%The power spectrum of the stellar atmosphere and that of the fringing signal have significant overlap, 

We find that it is much more difficult to fit the spectrum of an M-giant star compared to bright, A0V/F0V stars because of mismatch between the stellar model and the data and the commensurate nature of the power spectrum of the fringing signal and the stellar lines. The amplitude of the residuals due to the mismatch is the same order of magnitude as the fringing residuals, making it difficult to avoid fitting model mismatch while fitting the fringing signal. No best fit parameters are reported because of this. Better stellar models are necessary to address fringing found in M-giant stellar spectra.

\section{Wedging dichroics to remove unwanted fringes} 
\label{sec:wedge}
Since the parallelism of the dichroics creates several Fabry-P\'{e}rot cavities, we decided to modify the optics to alter the optical path of the co-propagating ghost, preventing it from re-entering the beam or falling on the detector. To combat the most difficult to parameterize and characterize fringing signals attributed to the dichroics, we introduced deviation by wedging the dichroics. We ordered new dichroics with a wedge angle of 50$''$, keeping all other properties consistent with the previous, unwedged dichroics. At 2 \textmu m, this results in approximately a 5 $\lambda$/D offset between the primary beam and the co-propagating ghost, as shown in \autoref{fig:fab_diagram}. The new dichroics were installed in April 2024 as part of a KPIC service mission \cite{Echeverri2024, Jovanovic2025}.

\begin{figure}
  \centering
  \includegraphics[trim={0cm 0cm 0cm 0cm},clip,width=\linewidth]{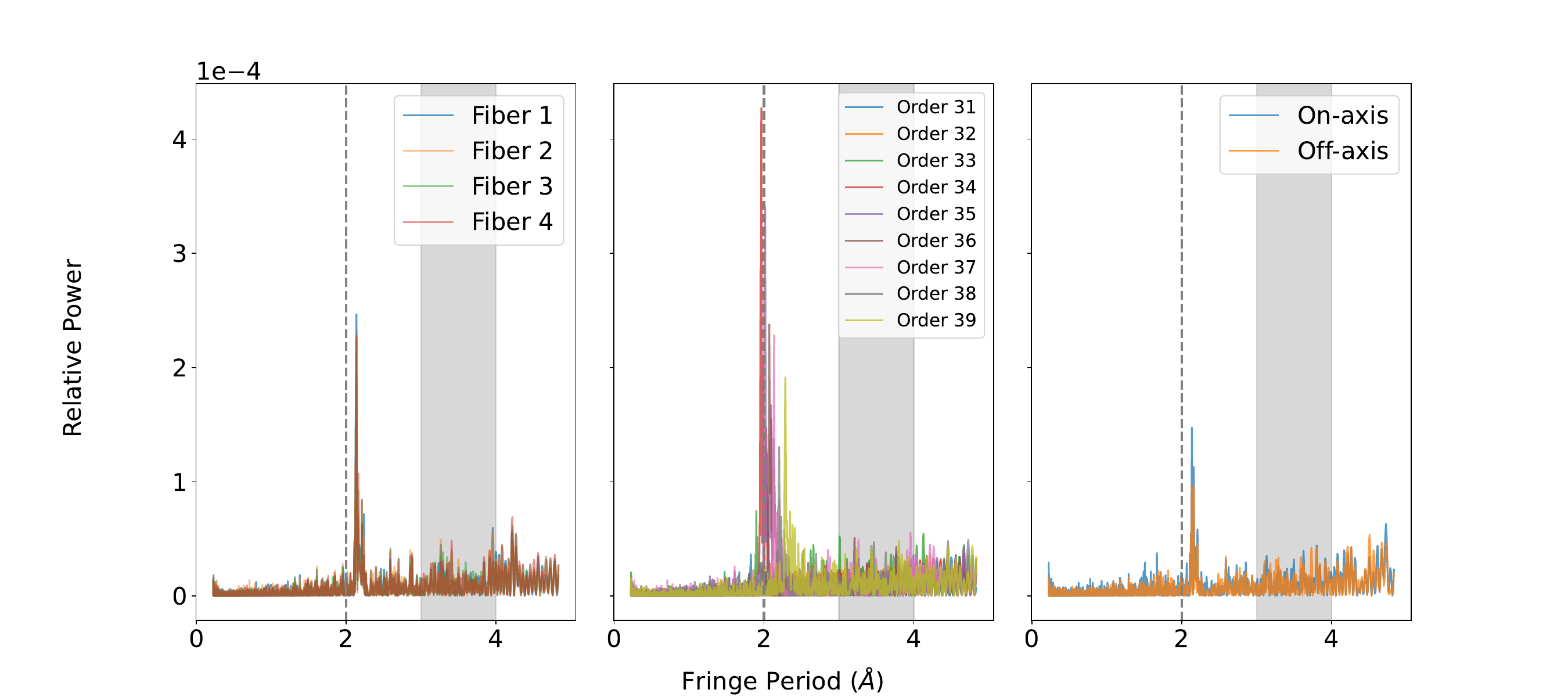}
  \caption{Lomb-scargle periodogram of the residuals (data-model) to confirm the suppression of the fringing signal between 3-4 \text{\AA}. \textbf{Left: }The Lomb-scargle periodogram of the residuals (data-model) of HIP 61960 for each KPIC fiber. \textbf{Middle: }The Lomb-scargle periodogram of the residuals (data-model) of HIP 61960 for each KPIC science order. \textbf{Right: }The Lomb-scargle periodogram of the residuals (data-model) of on-axis and off-axis observations of HIP 52686. In each instance, the fringing signal from the NIRSPEC entrance window at 2\text{\AA} is present while the fringing signal from the dichroics between 3-4 \text{\AA} is mitigated. The gray dashed line denotes the period of the fringing signal due to the NIRSPEC entrance window, while the gray shaded region denotes the period range of fringing signal due to unwedged dichroics.}
\label{fig:wedge_fringes}
\end{figure}

After wedging the dichroics in April of 2024, we confirm the presence of fringes due to the KPIC dichroics are no longer apparent. \autoref{fig:wedge_fringes} shows the Lomb-scargle periodogram of the residuals from our forward model fits for each KPIC fiber. As expected, we still observe the fringing signal at 2\text{\AA} due the NIRSPEC entrance window, but no longer observe fringes between 3-4 \text{\AA} due to the dichroics. We also verify that the dichroic fringing signal is no longer observed in different spectral orders or when offsetting to a companion. Although wedging the dichroics was effective in mitigating the 3-4 \text{\AA} fringing signal, it does not remove fringing found in archival KPIC data, motivating the usefulness of modeling the signal in \autoref{sec:solutions}.

\subsection{Application to A0V calibration star after wedging dichroics} 
\label{sec:A0_wedge}
\begin{figure*}
  \centering
  \includegraphics[width=\linewidth]{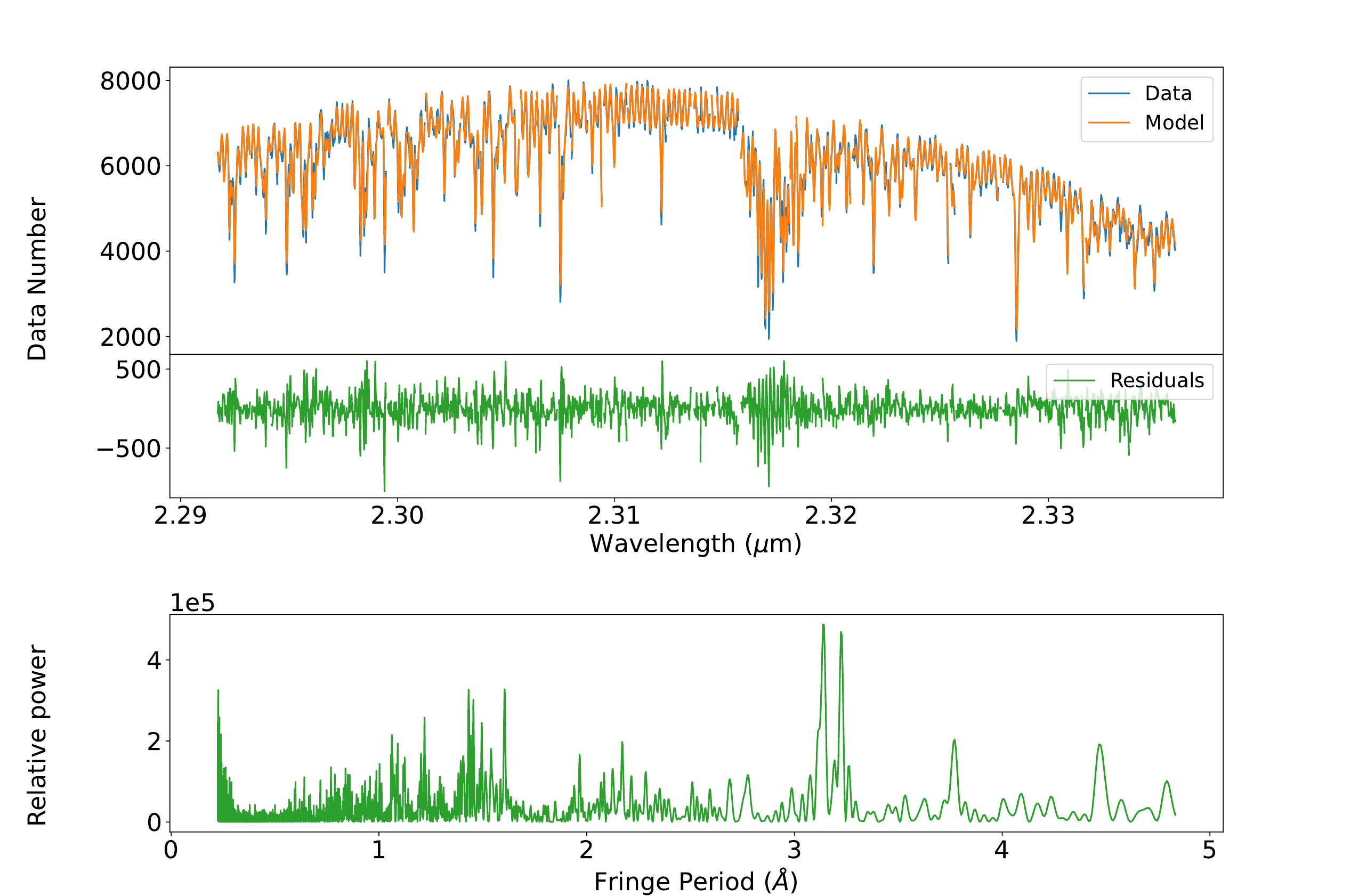}
  \caption{Example of forward modeling of the KPIC fringing signal, after wedging the dichroics, on a telluric calibrator star. \textbf{Top: }Spectrum of HIP 61960 after incorporating a fringing signal in the forward model framework. The data is in blue, the model is in orange, and the residuals (data-model) are in green. \textbf{Bottom:} A Lomb-scargle periodogram of the residuals of the model including the fringing signal. The incorporation of the the fringing model reduces the amplitude of the residuals by a factor of 10 compared to the model without the fringing model. The signal is suppressed for fringing due to the NIRSPEC entrance window at 2\text{\AA}, but an additional signal is seen at 3-4 \text{\AA} that is not apparent in \autoref{fig:wedge_fringes}. The excess power between 3-4 \text{\AA} could be due to leftover ghosts from the wedged dichroics, systematic mismatch in the line profiles of telluric model fits to the data, inaccurate flux extraction, or other currently unidentified systematics.}
  \label{fig:wedge_fringes_spec}
\end{figure*}

Since two of three sources of fringing are mitigated due to wedging the dichroics, we once again apply our model in \autoref{eq:fringe_model} to on-sky observations with the new optics installed. However, since we now only see a single Fabry-P\'{e}rot cavity, we fit for 3 parameters instead of the original 8. \autoref{fig:wedge_fringes_spec} shows the fit of the forward model incorporating the fringing signal while the best fit parameters can be found in \autoref{table:fringe_fit}.

From modeling the fringing, the signal is suppressed at the characteristic period of 2\text{\AA} for the NIRSPEC entrance window and the amplitude of the residuals is reduced by a factor of 10, pushing KPIC observations closer to the photon noise limit than ever before. However, an additional signal is seen at 3-4 \text{\AA} that is not apparent in \autoref{fig:wedge_fringes} and is difficult to notice except in analysis of residuals. The excess power between 3-4 \text{\AA} could be due to leftover ghosts from the wedged dichroics, systematic mismatch in the telluric model fits to the data due to approximations of molecular collisional physics near atmospheric pressures, inaccurate flux extraction, or other currently unidentified systematics as explained in more detail in Wang et al.\cite{Wang2024}.

\section{Conclusions and future work} 
\label{sec:future}
In  this paper, we identify three unique sources of fringing due to transmissive optics in KPIC/NIRSPEC and adopt a physically-informed model of a Fabry-P\'{e}rot cavities to effectively model the fringing signal seen in on-axis observations of A0V/F0V stars. We wedge two of the transmissive optics internal to KPIC and verify the absence of the fringing signal due to the unwedged dichroics. By applying our fringing model to observations of A0V stars where the wedged dichroics are present, we find the fringing signal is suppressed at the characteristic period of 2\text{\AA} for the NIRSPEC entrance window and the amplitude of the residuals is reduced by a factor of 10. However, wedging the transmissive optics for future KPIC operations does not address fringing in archival data or account for fringing due to the NIRSPEC entrance window, so the modeling framework for postprocessing data is still necessary. Understanding fringing and how to model it effectively continues to be relevant since transmissive optics are used frequently in the design of astronomical instruments. Generalizing the results of these findings will help others understand and suppress fringing signals within current and future instruments. 

In the future, we need to verify that the NIRSPEC entrance window fringing signal is truly stable by looking at observations of the same star on different nights, across different fibers, between different orders, and on and off axis. Once we determine the entrance window is stable, we can probe whether the physically informed model can be used on all observations. Additionally, since there is only one prominent source of fringing after wedging the KPIC dichroics, which can be described by only three parameters, we plan to re-explore applying the forward model of the fringing to M-giant wavelength calibration stars to improve the wavelength solution of KPIC observations. We will also extend these findings to the High-resolution Infrared Spectrograph for Exoplanet Characterization, Keck/HISPEC (R$\sim$100,000 between $\text{0.98 - 2.50} \mu$m) \cite{Konopacky2023}, which has an expected first light date in late 2026. 

\section{Code and data availability statement} 
The data that supports the findings of this article are not publicly available. They can be requested from the author at khorstma@astro.caltech.edu.

\section{Disclosures}
The authors declare there are no financial interests, commercial affiliations, or other potential conflicts of interest that have influenced the objectivity of this research or the writing of this paper.

\acknowledgments 
K.H. is supported by the National Science Foundation Graduate Research Fellowship Program under Grant No. 2139433. J.X. is supported by the NASA Future Investigators in NASA Earth and Space Science and Technology (FINESST) award \#80NSSC23K1434. Funding for KPIC has been provided by the California Institute of Technology, the Jet Propulsion Laboratory, the Heising-Simons Foundation (grants \#2015-129, \#2017-318, \#2019-1312, \#2023-4598), the Simons Foundation, and the NSF under grant AST-1611623. An earlier version of this work was presented in the Proceedings of SPIE \cite{Horstman2024SPIE}.

The W. M. Keck Observatory is operated as a scientific partnership among the California Institute of Technology, the University of California, and NASA. The Keck Observatory was made possible by the generous financial support of the W. M. Keck Foundation. We also wish to recognize the very important cultural role and reverence that the summit of Maunakea has always had within the indigenous Hawaiian community. We are most fortunate to have the opportunity to conduct observations from this mountain and K.H. wishes to acknowledge that the astronomical observations in this paper were only possible because of the dispossession of Maunakea from the Kan\={a}ka Maoli.

% References
\bibliography{main} 
\bibliographystyle{spiejour} 

\end{document}